\documentclass[11pt,oneside,a4paper]{article}
\usepackage[T1]{fontenc}
\usepackage{newtxtext,newtxmath}

\let\OLDthebibliography\thebibliography
\renewcommand\thebibliography[1]{
  \footnotesize
  \OLDthebibliography{#1}
  \setlength{\parskip}{0pt}
  \setlength{\itemsep}{0pt plus 0.3ex}
}

\begin{document}
\pagestyle{plain}

{\centering
  {
    {\bf \Large Rare decays at the CERN high-intensity kaon beam facility
    }\\[3ex]
    \large The NA62/KLEVER Collaborations\\[0.25ex]
  }\par
}

\vspace*{3ex}

Precision measurements of the branching ratios (BRs) for rare kaon decays
can provide unique constraints on CKM unitarity and may reveal the
existence of new physics.
The BRs for two of these decays, $K^+\to\pi^+\nu\bar{\nu}$ and
$K_L\to\pi^0\nu\bar{\nu}$, 
are strongly suppressed in the Standard
Model (SM), while their rates can be calculated with very small
theoretical uncertainties~\cite{Buras:2015qea}.
These decays are potentially sensitive to
mass scales of hundreds of TeV~\cite{Buras:2014zga}
and several models of new physics predict large
deviations from the SM~\cite{Chen:2018ytc,Bobeth:2017ecx,Bobeth:2016llm,Endo:2016tnu,Endo:2017ums,Crivellin:2017prd,Blanke:2015wba}.
While evidence for lepton-flavor-universality (LFU) violation in $B$
decays is mounting
~\cite{Aaij:2017vbb,Aaij:2019bzx,Aaij:2019wad,Aaij:2020nrf},
most explanations
predict strong third-generation couplings and thus significant changes
to the $K\to\pi\nu\bar{\nu}$ BRs through couplings to final states with
$\tau$ neutrinos~\cite{Bordone:2017lsy}.
Measurements of the $K\to\pi\nu\bar{\nu}$ BRs 
may demonstrate that LFU violation
is a manifestation of new degrees of freedom 
such as
leptoquarks~\cite{Buttazzo:2017ixm,Fajfer:2018bfj}.
Anomalously high apparent values for the $K\to\pi\nu\bar{\nu}$ BRs might
also be interpreted as evidence for the existence of hidden-sector particles,
such as a dark scalar $X$ produced in place of the $\nu\bar{\nu}$
pair~\cite{Egana-Ugrinovic:2019wzj}. 
Because the $K^+$ and $K_L$ decays have different sensitivity to new
sources of $CP$ violation, measurements of both BRs would be useful not only
to uncover evidence of new physics in the quark-flavor sector but also
to shed light on its nature.

Much progress has been made recently in the measurement of the
$K\to\pi\nu\bar{\nu}$ BRs.
The KOTO experiment at J-PARC~\cite{Shimizu:2020xxx} has
achieved a preliminary single-event sensitivity (SES) for $K_L\to\pi^0\nu\bar{\nu}$ decays
of $7.1\times10^{-10}$, corresponding to 0.04 SM signal events
collected in 2016--2018. The collaboration has observed 3 candidate events,
with an estimated background of $1.05\pm0.28$ events, dominantly from
$K^+$ charge-exchange, techniques for the suppression of which are under development. KOTO expects to reach the SES for the SM decay
by about 2025. A possible KOTO Step-2 upgrade thereafter would require
construction of a planned extension of the J-PARC hadron hall, as well as
of a completely new detector several times larger than the present
detector~\cite{Nomura:2019xxx}.
Meanwhile, the NA62 experiment at the CERN SPS~\cite{NA62:2017rwk} has
announced preliminary results from 2018 data taking: 17 candidate events
for the $K^+\to\pi^+\nu\bar{\nu}$ decay were observed, with an 
expected SM signal of $7.6\pm 0.8$ and an 
expected background of $5.3^{+1.0}_{-0.7}$ events~\cite{Marchevski:2020xxx}.
In combination with the smaller
data set collected in 2016-2017 (three candidate events)~\cite{CortinaGil:2020vlo, CortinaGil:2018fkc}, NA62 obtains
${\rm BR}(K^+\to\pi^+\nu\bar{\nu}) = (11.0^{+4.0}_{-3.5}\pm0.3)\times 10^{-11}$,
improving on the previous bounds from the Brookhaven E787
and E949 experiments~\cite{Artamonov:2009sz} and giving the first significant evidence ($>3\sigma$) for this rare decay. NA62 expects to attain
$\sim$10\% precision on the BR with data to be collected in 2021--2024.

Following upon NA62's successful application of the in-flight technique
to measure ${\rm BR}(K^+\to\pi^+\nu\bar{\nu})$,
we envision a comprehensive program for
the study of the rare decay modes of both $K^+$ and $K_L$
mesons, to be carried out with high-intensity kaon beams from the
CERN SPS in multiple phases, including both an experiment to measure
${\rm BR}(K^+\to\pi^+\nu\bar{\nu})$ at the 5\% level and an experiment
to measure ${\rm BR}(K_L\to\pi^0\nu\bar{\nu})$ at the 20\% level, which together with the KOTO measurement will push the precision below 15\%.
The detectors could also be reconfigured to allow measurements of
$K_L$ decays with charged particles, such as $K_L\to\pi^0\ell^+\ell^-$.

The success of this program depends on the delivery of
a high-intensity, slow-extracted 400-GeV/$c$ proton beam from the SPS
to the current NA62 experimental hall, which is ideally suited for
next-generation kaon experiments.
Up to $10^{19}$ protons on target per year will be required.
Studies indicate that, with duty-cycle optimization,
sufficient proton fluxes can be delivered to allow a high-intensity
kaon experiment to run as part of a robust fixed-target
program~\cite{Bartosik:2018xxx}.
The 
upgrades to the primary beamline tunnel, target gallery,
and experimental cavern to handle an intensity of up to six times that in
NA62 has been studied
as part of the Physics Beyond Colliders initiative at CERN
~\cite{Banerjee:2018xxx,Gatignon:2018xxx}. 

To measure ${\rm BR}(K^+\to\pi^+\nu\bar{\nu})$ with 5\% precision,
the experiment must be able to handle a beam
intensity of four times that in NA62. The entire complement of
NA62 detectors~\cite{NA62:2017rwk} would require substantial upgrades in order to provide
clean event reconstruction at the expected rates. 
Development of the new detectors is synergistic with 
R\&D efforts for HL-LHC.
Ultrafast ($\sigma_t \sim 20$~ps) single-photon detectors,
such as advanced devices making use of multi-channel plates, will be needed
for use in the Cherenkov detectors for beam and secondary particle
identification. 
Tracking of the $K^+$ beam will require a new Gigatracker with 50~ps
time resolution, capable of operating at
rates of up to 8 MHz/mm$^2$; possible solutions under exploration include
silicon sensors incorporating LGAD technology~\cite{Cartiglia:2020eaf}
and 3D pixels with trench geometry~\cite{Mendicino:2019ern}.
Design work has started on a tracking system for secondary particles based
on thin-walled ($\sim20~\mu$m), narrow-gauge (5 mm) straw tubes operating
in vacuum, with a time resolution of $\sim$6 ns per straw. This work, which
builds on the NA62 straw-tracker design, is synergistic with R\&D work for
COMET phase-II at J-PARC~\cite{Nishiguchi:2017gei}.

The baseline design for the $K_L$ experiment is the KLEVER
project~\cite{Ambrosino:2019qvz},
presented as part of the Physics Beyond Colliders
initiative~\cite{Beacham:2019nyx}. The KLEVER goal is to measure
${\rm BR}(K_L \to\pi^0\nu\bar{\nu})$ to within 20\%.
The measurement technique is complementary to that for KOTO: the boost from the high-energy neutral beam facilitates the rejection of
background channels such as $K_L\to\pi^0\pi^0$ by detection of the
additional photons in the final state.
Background from $\Lambda \to n\pi^0$ decays in fiducial volume must
be kept under control, and recent work has focused on the possibility
of a beamline extension and other adaptations of the experiment to ensure
sufficient rejection of this channel.  
The layout poses challenges for the design of the small-angle
vetoes, which must reject photons from $K_L$ decays escaping through the
beam exit amidst an intense background from soft photons and neutrons in the
beam. A notable feature of KLEVER is the exploitation of coherent interaction
effects in oriented
crystals~\cite{Bak:1988bq,Kimball:1985np,Baryshevsky:1989wm}
to remove high-energy photons from the
neutral beam and, possibly, for the construction of small-angle
photon vetoes with high conversion efficiency for photons but good
transparency to beam neutrons.

Experiments with different kaon beams would run
consecutively
in the same experimental area
and use interchangeable detectors.
The forward calorimetry and photon veto systems would be 
the same for both
experiments.
Detector and beamline configuration between phases will be scheduled during an LHC shutdown period.
As in NA62, the physics program of the
high-intensity $K^+$ experiment is broad and includes studies of radiative
kaon decays, 
tests of lepton-universality conservation,
searches for lepton-flavor or lepton-number violating decays, decays of the kaon to hidden-sector particles~\cite{snowmassdark}, 
and other precision tests. 
In the $K_L\to\pi^0\nu\bar{\nu}$ configuration, to optimize the
measurement,
the $K_L$ experiment will have no secondary tracking and limited PID
capability.
As part of a broader physics program, however,
we foresee the possibility of taking data with the downstream detectors
configured for charged-particle measurements (calorimetry, secondary
tracking and PID) but with a neutral beam. This will allow searches for
radiative and lepton-flavor violating $K_L$ decays, $K_L$ decays to exotic
particles, and particularly,
measurements of the $K_L\to\pi^0\ell^+\ell^-$ BR, for which
only limits exist at present~\cite{AlaviHarati:2003mr,AlaviHarati:2000hs}. 
Measurement of these BRs would overconstrain
the kaon unitarity triangle; to the extent that there is little constraint
on the CP-violating phase of the $s\to d\ell^+\ell^-$ transition, measurement
of the $K_L \to\pi^0\ell^+\ell^-$ BRs may reveal the effects of new physics
\cite{Smith:2014mla}.
The combination of data from multiple channels will yield a thorough investigation of new physics in the kaon sector. 

The program described builds on a CERN tradition of groundbreaking
experiments in kaon physics, following on the successes of NA31, NA48, NA48/1,
NA48/2, and NA62.
In the 2020 update of the European Strategy for Particle Physics,
rare kaon decay measurements at CERN were identified as contributing to a
diverse program of high-impact particle physics initiatives that is
an essential part of the Strategy~\cite{CERN:2020xxx}.

\newpage
\bibliographystyle{elsarticle-num-mod}
\bibliography{snowmass}

\end{document}